\documentclass[fleqn,twoside]{article}
\usepackage{amsmath,amssymb,graphicx,espcrc2}

\title{Gluon field distribution in baryons} 

\author{F. Bissey\address[PN]{Institute of Fundamental Sciences, Massey University,\\
        Private Bag 11 222, Palmerston Noth, New Zealand}, 
        F-G. Cao\addressmark, A. Kitson\addressmark, 
        B. G. Lasscock\address[CSSM]{Centre for the Subatomic Structure of Matter,\\
        Physics Building, University of Adelaide, SA 5005, Australia},
        D. B. Leinweber\addressmark[CSSM],
        A. I. Signal\addressmark[PN], 
        A. G. Williams\addressmark[CSSM] and 
        J. M. Zanotti\addressmark[CSSM]\address{John von
        Neumann-Institut f{\"u}r Computing NIC,\\ 
        Deutsches Elektronen-Synchrotron DESY, D-15738 Zeuthen, Germany}}

\begin{document}

\begin{abstract}
Methods for revealing the distribution of gluon fields within the
three-quark static-baryon potential are presented.  In particular, we
outline methods for studying the sensitivity of the source on the
emerging vacuum response for the three-quark system.  At the same
time, we explore the possibility of revealing gluon-field
distributions in three-quark systems in QCD without the use of
gauge-dependent smoothing techniques.  Renderings of flux tubes from a
preliminary high-statistics study on a $12^3 \times 24$ lattice are
presented.
\end{abstract}

\maketitle

\section{INTRODUCTION}

Recently there has been renewed interest in studying the distribution
of quark and gluon fields in the three-quark static-baryon potential.
While early studies were inconclusive \cite{Flo86}, improved computing
resources and analysis techniques now make it possible to study this
system in a quantitative manner \cite{Tak01,Ale02}.  In particular, it
is possible to directly compute the gluon flux distribution
\cite{Ich02,Oki03} using lattice QCD techniques similar to those
pioneered in mesonic static-quark systems \cite{Som86,Bal95,Hay96}.

Like Okiharu and Woloshyn \cite{Oki03} our first interest is to test
the static-quark source-shape dependence of the observed flux
distribution, as represented by correlations between the quark
positions and the action or topological charge density of the gauge
fields.  To this end, we choose three different ways of connecting the
gauge link paths required to create gauge-invariant Wilson loops.  In
the first case, quarks are connected along a T-shape path, while in
the second case an L-shape is considered.  Finally symmetric link
paths approximating a Y-shape such that the quarks approximate an
equilateral triangle are considered.  The latter is particularly
interesting as the probability of observing $\Delta$-shape flux-tubes
are maximized in this equidistant case.

Because the signal decreases in an exponential fashion with the size
of the loop, it is essential to use a method to enhance the signal.
We use strict three-dimensional APE smearing \cite{APE87} to enhance
the overlap of our spatial-link paths with the ground-state
static-quark baryon potential.  Time-oriented links remain untouched
to preserve the correct static quark potential at all separations.

Excellent signal to noise is achieved via a high-statistics approach
based on translational symmetry of the four-dimensional lattice volume
and rotational symmetries of the lattice as described in detail below.
This approach contrasts previous investigations using gauge fixing
followed by projection to smooth the links and resolve a signal in the
flux distribution \cite{Ich02,Oki03}.

\section{WILSON LOOPS}

To study the flux distribution in baryons on the lattice, one begins
with the standard approach of connecting static quark propagators by
spatial-link paths in a gauge invariant manner.  APE-smeared
spatial-link paths propagate the quarks from a common origin to their
spatial positions.  

The smearing procedure replaces a spatial link, $U_{\mu}(x)$, with a
sum of $1-\alpha$ times the original link plus $\alpha/4$ times its
four spatially oriented staples, followed by projection back to
$SU(3)$.  We select the unitary matrix $U_{\mu}^{\rm FL}$ which
maximizes $ {\cal R}e \, {\rm{tr}}(U_{\mu}^{\rm FL}\,
U_{\mu}'^{\dagger})$, where $U_{\mu}'$ is the smeared link, by
iterating over the three diagonal $SU(2)$ subgroups of $SU(3)$
repeatedly.  We repeat the combined procedure of smearing and
projection $10$ times, with $\alpha = 0.7$.

Untouched links in the time direction propagate the spatially
separated quarks through Euclidean time.  For sufficient time
evolution the ground state is isolated.  Finally smeared-link
spatial paths propagate the quarks back to the common spatial
origin.

The three-quark Wilson loop is defined as:
\begin{equation}
W_{3Q}=\frac{1}{3!}\epsilon^{abc}\epsilon^{a'b'c'} \, U_1^{aa'} \,
U_2^{bb'} \, U_3^{cc'},
\end{equation}  
where $U_j$ is a staple made of path-ordered link variables
\begin{equation}
U_j \equiv P \exp \left( ig \int_{\Sigma_j} dx_{\mu} \, A^{\mu}(x)
\right) \, ,
\end{equation}
and $\Sigma_j$ is the path along a given staple.  

In this study, we consider two-dimensional spatial-link paths. Figures
1 and 2 give the projection of T and L shape paths in the $xy$ plane.
Figure 3 shows the idea behind the Y-shape.  In all cases the three
quarks are created at the origin, O (white bubble), then are
propagated to the positions Q1, Q2 or Q3 (black circle) before being
propagated through time and finally back to a sink at the same spatial
location as the source (O).

\begin{figure}[t]
\centering\includegraphics[width=3cm,height=2cm,clip=true]{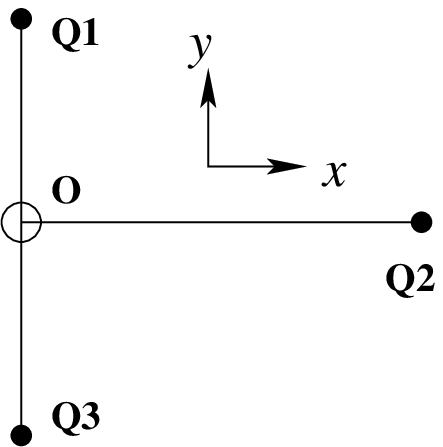}
\caption{Projection of the T-shape path on the $xy$-plane.}
%
\centering\includegraphics[width=3cm,height=2cm,clip=true]{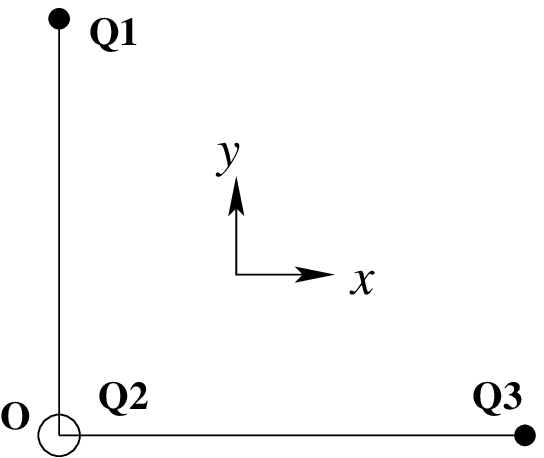}
\caption{Projection of the L-shape path on the $xy$-plane.}
\centering\includegraphics[width=3cm,height=2cm,clip=true]{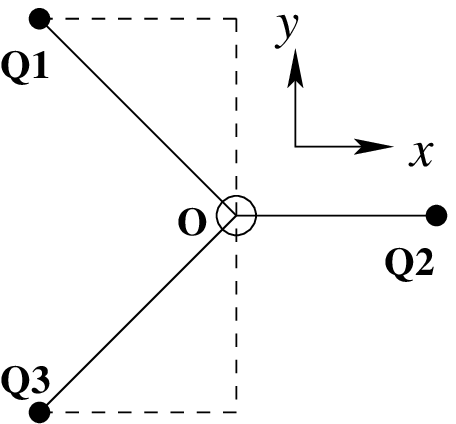}
\caption{Projection of the Y-shape path on the $xy$-plane.}
\end{figure}

For the Y-shape we create elementary diagonal ``links'' in the form of
boxes as shown in Fig. 4.  The $1 \times 1$ and $1 \times 2$ boxes are
the average of the two path-ordered link variables going from one
corner to the diagonally opposite one. Taking both of these paths
better maintains the symmetry of the ground state potential and
therefore provides improved overlap with the ground state.  We also
consider $2 \times 3$ boxes which are the averages of the possible
paths connecting two opposite corners using 1x1 and 1x2 boxes.  We
will create further link paths in the future for use in bigger loops
as necessary.  Hence a diagonal staple is, in fact, an average of
several ``squared path'' staples connecting the same end points.  

\begin{figure}[t]
\centering\includegraphics[height=1.5cm,clip=true]{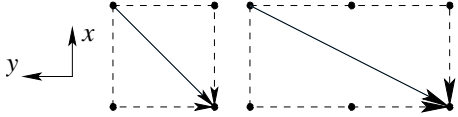}
\caption{The diagonal path (solid line) is taken as the average of
the two possible ones going around the box enclosing the end points.}
\end{figure}

The quark co-ordinates considered for Y-shape paths are summarized in
Table~\ref{tab:coord}.  We note that the origin $(0,0)$ is not at the
centre of these coordinates.  Rather the coordinates are selected to
place the quarks at approximately equal distances from each other.

\begin{table}[t]
\centering
\begin{tabular}{ccccc}
\hline
\noalign{\smallskip}
\multicolumn{3}{c}{$(x,y)$ Coordinates}
         &\multicolumn{2}{c}{Separation} \\
$Q_1$  &$Q_2$  &$Q_3$  &Cart. &Diagonal \\
\hline
\hline
\noalign{\smallskip}
$( 0, 1)$ &$( 1, -1)$ &$( -1, -1)$  &2         &2.24 \\
$( 0, 2)$ &$( 2, -1)$ &$( -2, -1)$  &4         &3.61 \\
$( 0, 3)$ &$( 2, -1)$ &$( -2, -1)$  &4         &4.47 \\
$( 0, 3)$ &$( 3, -2)$ &$( -3, -2)$  &6         &5.83 \\
$( 0, 4)$ &$( 4, -3)$ &$( -4, -3)$  &8         &8.06 \\
$( 0, 5)$ &$( 5, -4)$ &$( -5, -4)$  &10        &10.3 \\
\hline
\end{tabular}
\caption{ $(x,y)$ coordinates for the three quarks considered for
  Y-shape paths.  The separation of the two quarks along a Cartesian
  (Cart.)  direction are compared to the separation of those from the
  third quark.  }
\label{tab:coord}
\end{table}

\section{GLUON FIELD CORRELATION}

In this investigation we characterize the gluon field by the action
density $S(\vec y, t)$ observed at spatial coordinate $\vec y$ at
Euclidean time $t$ measured relative to the origin of the three-quark
Wilson loop.  We calculate the action density using the
highly-improved ${\cal O}(a^4)$ three-loop improved lattice
field-strength tensor \cite{Bilson-Thompson:2002jk} on four-sweep
APE-smeared gauge links.

Defining the quark positions as $\vec r_1$, $\vec r_2$
and $\vec r_3$ relative to the origin of the three-quark Wilson loop,
and denoting the Euclidean time extent of the loop by $T$, we evaluate
the following correlation function
\begin{eqnarray}
\lefteqn{C(\vec y; \vec r_1, \vec r_2, \vec r_3; T) =} \nonumber \\
&&\quad \frac{
\bigl\langle W_{3Q}(\vec r_1, \vec r_2, \vec r_3; T) \,
             S(\vec y, T/2) \bigr\rangle }
{
\bigl\langle  W_{3Q}(\vec r_1, \vec r_2, \vec r_3; T) \bigr\rangle \,
\bigl\langle S(\vec y, T/2) \bigr\rangle
}
  \, ,
\label{correl}
\end{eqnarray} 
where $\langle \cdots \rangle$ denotes averaging over configurations
and lattice symmetries as described below.
This formula correlates the quark positions via the three-quark Wilson
loop with the gauge-field action in a gauge invariant manner.  For
fixed quark positions and Euclidean time, $C$ is a scalar field in
three dimensions.  For values of $\vec y$ well away from the quark
positions $\vec r_i$, there are no correlations and $C \to 1$.

This measure has the advantage of being positive definite, eliminating
any sign ambiguity on whether vacuum field fluctuations are enhanced
or suppressed in the presence of static quarks.  We find that $C$ is
generally less than 1, signaling the expulsion of vacuum fluctuations
from the interior of heavy-quark hadrons.

\section{STATISTICS}

For this work we consider 200 quenched QCD gauge-fields using the
${\cal O}(a^2)$-mean-field improved Luscher-Weisz plaquette plus
rectangle gauge action \cite{Luscher:1984xn} on $12^3\times 24$
lattices at $\beta=4.60$, providing a lattice spacing of $a =
0.123(2)$~fm as set by the string tension.

To improve the statistics of the simulation we use various symmetries
of the lattice.  First, we make use of translational invariance by
computing the correlation on every node of the lattice, averaging the
results over the four-volume.

To further improve the statistics, we use reflection symmetries.
Through reflection on the plane $x=0$ we can double the number of T
and Y-shaped Wilson loops. By using reflections on both the plane
$x=0$ and $y=0$ we can quadruple the number of L-shaped loops.

We finally use $90^\circ$ rotational symmetry about the $x$-axis to
double the number of Wilson loops. This means we are using both the
$xy$ and $xz$ planes as the planes containing the quarks.

For this exploratory study we present images for $T=2$ and note that
qualitatively similar results are observed for $T=4$.  We plan to
examine this and other similarly conservative alternatives on our
anticipated larger lattice volumes providing further statistics
improvement.

\section{SIMULATION RESULTS}

\begin{figure}[t]
\centering
\includegraphics[width=\hsize,clip=true]{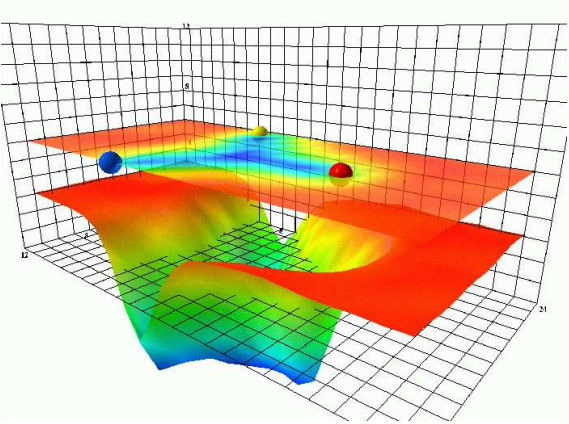}
\caption{Ortho-slice and associated surface plot of the correlation
  function of Eq.~(\protect\ref{correl}), $C(\vec y)$, observed for
  positions $\vec y$ in the plane of the three-quark system.
\label{ortho}}
\end{figure}

Figure \ref{ortho} plots the correlation function of
Eq.~(\ref{correl}), $C(\vec y)$, for an ortho-slice in the plane of
the three-quark system.  The colours of the ortho-slice, denoting the
scalar values of $C(\vec y)$, are rendered as a surface plot below.
Away from the quark positions $C(\vec y) \to 1$, but within the
three-quark system $C(\vec y) \sim 0.9$ indicating the suppression of
QCD vacuum field fluctuations from the region inside heavy-quark
hadrons.  Here the quarks are approximately 0.75 fm from the centre of
the distribution and are about 1.25 fm from each other.  

\begin{figure}[t]
\centering
\includegraphics[width=\hsize,clip=true]{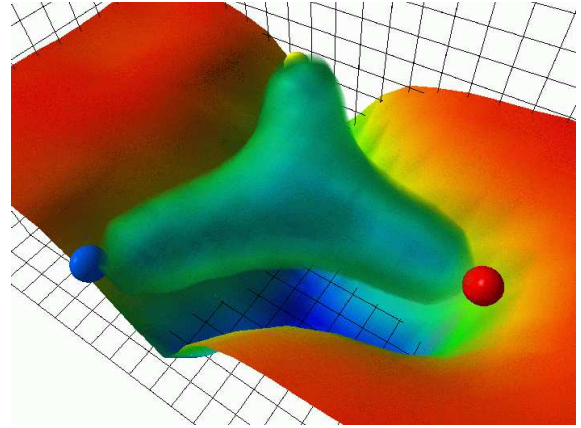}
\caption{Spatial points where vacuum field fluctuations are maximally
  suppressed in the three-quark system as measured by the correlation
  function, $C(\vec y)$, of Eq.~(\protect\ref{correl}).  The surface
  plot is as in Fig.~\protect\ref{ortho}.
\label{tubes}}
\end{figure}

Figure \ref{tubes} renders the spatial points of $C(\vec y)$ where the
gluon action density is largely suppressed.  Tube-like structures are
revealed in Y-shape as opposed to $\Delta$ shape in accord with
expectations from precision static quark-potential analyses.

We note that the centre of the flux tube is equidistant from the quark
positions and not centered over the origin of the Wilson loop.  This
gives us confidence that sensitivity to the source is minimal.
However it is important to examine this issue in greater detail and
this will be the subject of a forthcoming publication.

\end{document}